# Compilation, Optimization, Error Mitigation, and Machine Learning in Quantum Algorithms

Paul Wang, Jianzhou Mao, and Eric Sakk

Department of Computer Science, Morgan State University
Baltimore, Maryland 21251. USA

**Abstract.** This paper discusses the compilation, optimization, and error mitigation of quantum algorithms, essential steps to execute real-world quantum algorithms. Quantum algorithms running on a hybrid platform with QPU and CPU/GPU take advantage of existing high-performance computing power with quantum-enabled exponential speedups. The proposed approximate quantum Fourier transform (AQFT) for quantum algorithm optimization improves the circuit execution on top of an exponential speed-ups the quantum Fourier transform has provided.

## 1 Introduction

Quantum computing has advanced from computing using quantum chips alone to a quantum and High Performance Computing (HPC) hybrid approach with heterogeneous orchestration, development tools such as transpiler services, to abstraction services which focus on circuit and application functions. The Quantum Information Software Kit (*Qiskit*) Code Assistant uses Artificial Intelligence (AI) to help programmers generate function-related block code [1, 2].

Quantum algorithms are moving from classical pre-utility to error mitigation, while error correction is in sight. The new IBM 133 qubit Heron quantum chip is able to achieve 30% improvement in qubit coherence through active two-level system mitigation. Circuit depth can be significantly reduced by using the transpiled service with the help of AI [3]. With HPC workload manager, *Qiskit* has extended its capacity to support both quantum computers powered by Quantum Process Units (QPU) and high-performance classical computers powered with CPU, GPU, and Artificial Intelligence Unit (AIU) [4, 5].

The execution of a quantum algorithm is to 1) map problem instances to quantum circuits and operators, 2) optimize for target hardware execution, 3) execute via Runtime, a service and programming model for building, optimizing and executing quantum workloads, and 4) result processing or post-processing [6].

For mapping quantum computing problems to quantum circuits, Multi-Product Formulas (MPF) is able to reduce algorithmic errors by dividing the circuits into segmented and weighted circuits for executions [7]. Approximate Quantum Compiling (AQC)-tensor compresses the initial layers, allowing more circuit depth to be spent on further time evolution [8]. Sample-based Quantum Diagonalization (SQD) uses classical distributed computing to produce more accurate eigenvalue estimations from noisy quantum samples. It refines noisy samples with classical distributed computing to address large Hamiltonians [9].

Quantum Machine Learning (QML) uses the hybrid approach trains ML models on QPU, and runs the predictions on classical HPCs. With add-on tools, including MPF, AQC-Tensor, Operator Back Propagation (OBP), and SQD, speed-ups in chemistry simulation can be reached five times, the algorithm running speed from one year to 2.5 months [10, 11].

Parallelism and pipeline are emerging in quantum computing. Though it is not a pure classical Reduced Instruction Set Computer (RSIC) architecture with unified Cycles Per Instruction (CPI) concept, quantum pipelines are able to bring better speedups with the pipe stages increase [12–15].

Figure 1 shows the diagram of the quantum execution pipeline. To the left, (a) shows non-pipelined execution. To the right, (b) depicts the execution of a piplined quantum algorithm; note that at a given time, only one QPU is used. The pipeline improves the execution ≤ $n$ times, where n is the number of stages.

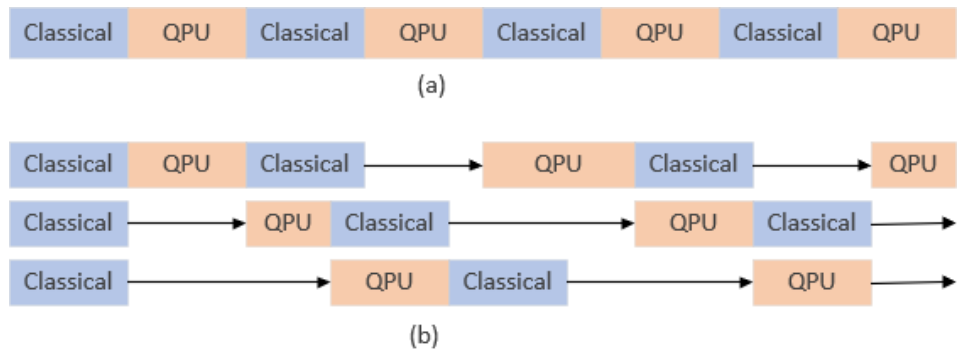

**Fig. 1.** Pipeline and Parallelism in Quantum Algorithm Execution.

## 2   Quantum Algorithm Transpilation and Optimization

Similar to compilers that convert high-level instructions to hardware-dependent machine language, transpilation is a process of converting a given quantum circuit to match the topology of a specific quantum chip and optimize the circuits for executions. Transpilers can run locally with Qiskits or remotely as a cloud service. The runtime service is critical for current resource-constraint quantum devices.

There are five stages for this pipeline:

- Init. Translate any gates that operate on more than two qubits into gates that only operate on one or two qubits.
- Layout. Lay each qubit to a specific quantum instruction set architecture (ISA).
- Translation. For a specific target ISA, translate the gates in a circuit to the native basis gates of a specified backend.
- Optimization. Decompose quantum circuits into the basis gate set of the target device, and try to reduce the depth from the layout and routing stages.
- Scheduling. Schedule tasks for circuits that have been translated to the target basis, mapped to the device, and optimized.

High-Order Binary Optimization (HOBO) can be solved using Q-CTRL's Qiskit function *FireOpal* Optimization Solver. for a problem to find the ground-state energy of a random-bond 156-qubit Ising model with Q-CTRL Solver compared to the classical random sampling, the probability increased from 0.04 to 0.1 while the cost reduced more than

200 times [16].

Circuit optimization involves rewriting the quantum circuit. Rewriting consists of six steps:

- Virtual circuit optimization
- 3+ qubit gate decomposition
- Placement on physical qubits
- Routing on restricted topology
- Translate to basis gates
- Physical circuit optimization

An example of circuit optimization is shown in Figure 2. To the left, a q0 and q2 swap followed by a cx q2 to q3 is equivalent to a cx q0 to q3 as shown on the right. For the same reason, a swap of q1 and q3 followed by a cx q3 to q1 (left) is equivalent to a cx q1 to q3 (right). The fact that a swap gate is not a native gate. It is an expensive operation to perform on noisy quantum devices that requires three CNOT gates. A Toffoli (ccx) gate is a three-qubit gate. The decomposition is quite costly that requires up to six CNOT gates and a number of single-qubit gates. Eliminating swap and Toffoli gates is a main goal in the transpilition process.

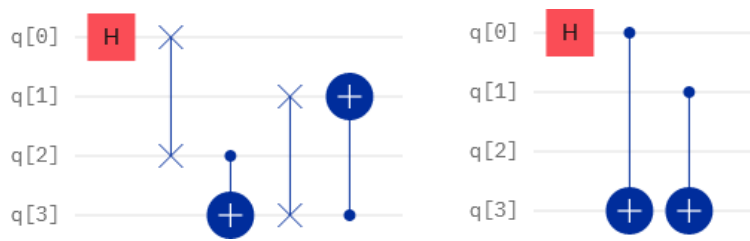

**Fig. 2.** Quantum Algorithm Tranpilation and Optimization

## 3  Quantum Error Mitigation and Suppression

There are many techniques for the mitigation and suppression of quantum errors. Dynamical decoupling is to insert pulse sequences into idle qubits to approximately cancel out the effect of coherent errors [17].

Pauli twirling is an error mitigation technique to compile random single-qubit gates into a logical circuit, achieving fault-tolerant quantum commutation [18].

Zero-noise extrapolation (ZNE) aims to mitigate error by estimating expectation values. It starts with amplification of the noise and then is followed by extrapolation [19]. Using the mirror circuit technique [20], which concatenates the circuit with its inverse circuit, the crosstalk error and infidelity of the quantum circuits can be revealed. Figure 3 shows a quantum circuit with random parameter and its mirror circuit.

## 4  Quantum Machine Learning

Qiskit comes with the Qiskit machine learning package that uses Quantum Neural Networks (QNNs) with two core primitives:

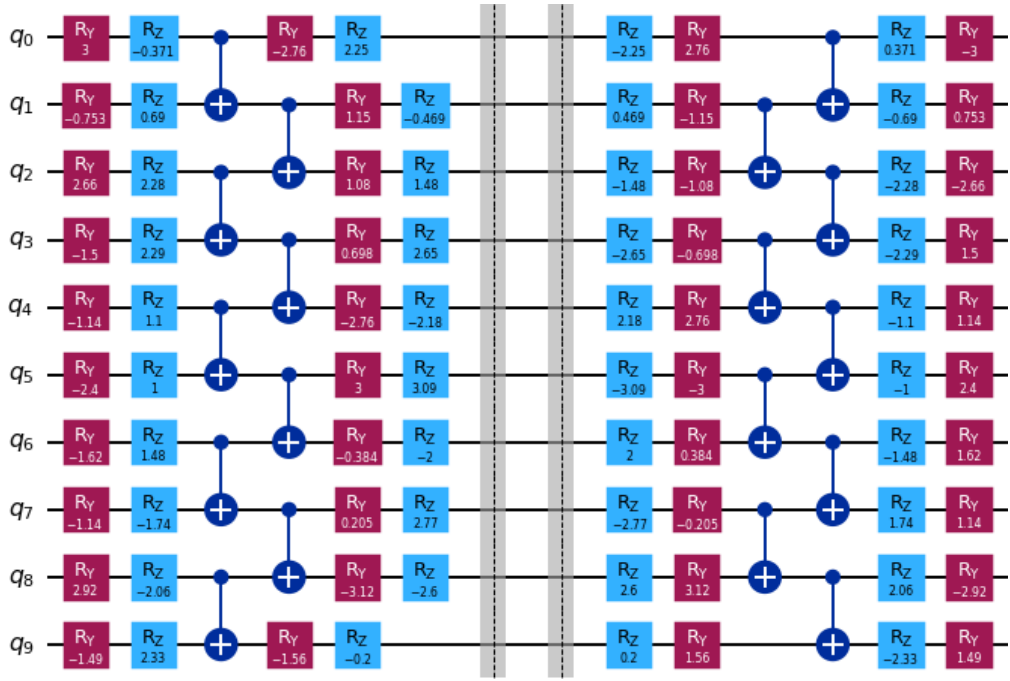

**Fig. 3.** Quantum Error Mitigation

- EstimatorQNN. Combine parametrized quantum circuits and Estimator primitive.
- SampleQNN. Translate bit-string into the desired outputs.

QML is integrated with PyTorch to allow easy integration into other packages [21, 22]. Figure 4 shows the classification results. It can be seen that class *0* is separated from other two classes, while overlaps exist in classes *1" and *2*. Data used for this plot were from the NASA Iris plants dataset.

## 5 Quantum Algorithm Optimization - AQFT

The order finding using quantum entanglement and modular exponentiation can be illustrated as follows:

1. Initialize two registers of qubits, first an argument register with t qubits and second a function register with $n = \lceil log_2 N \rceil$ bits. These registers start in the initial state: $|\psi_0\rangle = |0\rangle |0\rangle$.
2. Apply a Hadamard gate to form an equal weighted superposition of all integers: $|\psi_1\rangle = \frac{1}{\sqrt{T}} \sum_{a=0}^{T-1} |a\rangle |0\rangle$.
3. Implement modular exponentiation $x^a \bmod N$ on the function registers, the state becomes: $|\psi_2\rangle = \frac{1}{\sqrt{T}} \sum_{a=0}^{T-1} |a\rangle |x^a \bmod N\rangle$.
4. Perform a quantum Fourier transform on the argument register, resulting in the state: $|\psi_3\rangle = \frac{1}{T} \sum_{a=0}^{T-1} \sum_{z=0}^{T-1} e^{(2\pi)\left(\frac{aZ}{T}\right)} |Z\rangle |x^a \bmod N\rangle$.

Unlike using symbolic manipulation to solve problems, as we usually see in calculus, numerical algorithms use approximation to find solutions. This is especially useful in solving engineering problems, as well as solving problems in physics, life science, medicine, financial, and aerospace. Traditional error analysis on numerical algorithms focuses mostly on discretization errors. In other words, errors caused by using finite number of approximations of variables compared with continuous variables. Rearrange $|\psi_3\rangle$ and we get

$$|\psi_3\rangle = \frac{1}{\sqrt{T}} \sum_{z=0}^{T-1} \sum_{a_0} |x_0^a \mod N\rangle |Z\rangle \omega^{za_0} \sum_{n=0}^{T-1} e^{\left(\frac{2\pi i}{T}\right)znr}$$

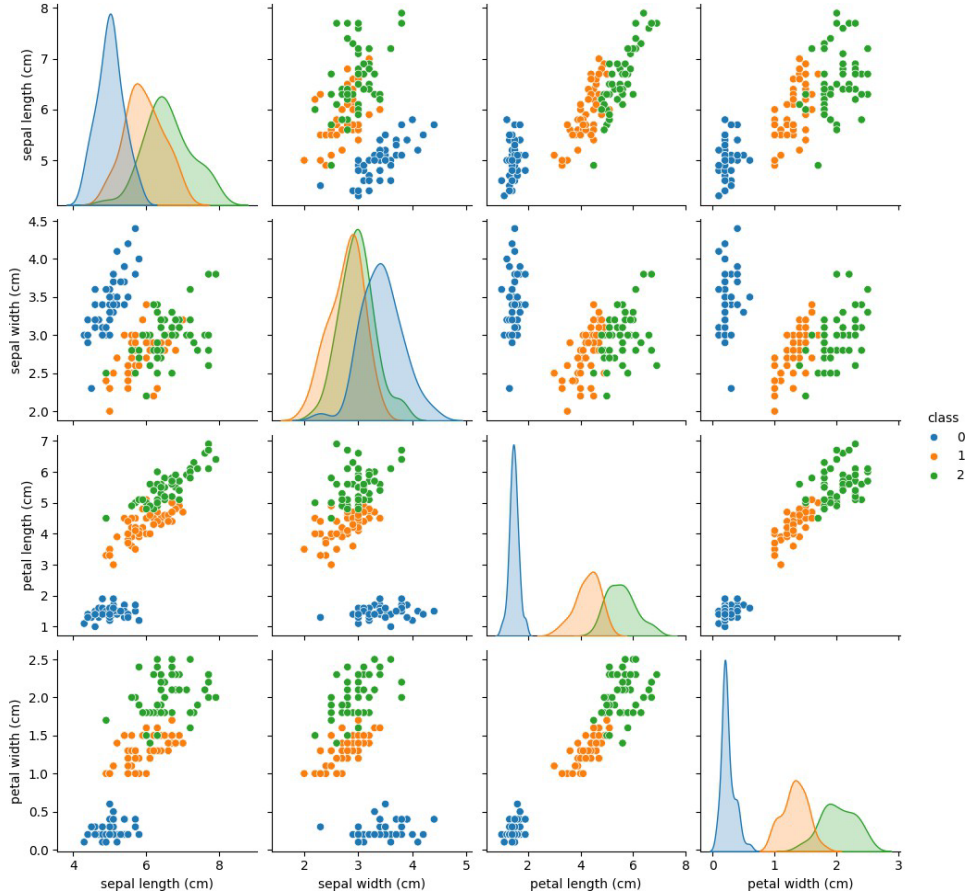

**Fig. 4.** QML Classification Result.

The phases spread to all directions, so most of them cancel each other out. For some cases where z is related to r through $a_0$, the phases are a multiple of $2\pi$. Figure 1 shows a QFT output with a series of "spikes" that occur when z is equal to $d$ time $T$ over $r$. This can be used to calculate the period $r$. Figure 5 shows the results of the quantum phase operation in the frequency domain.

## 5.1 Recursion Depth Analysis

The transforms on a single qubit [16] are done by recursively applying a Hadamard transform followed by rotations on condition $a_i$, here $e(0.a_k a_{k-1} \ldots a_1) = e^{a k/2^n}$. Note that $e^{a k/2^n}$ is a series of phase operations (ration along the z axis). Each time, the rotation angle reduces by half with an incremental increase of $k$.

Looking at the earlier equation, when $k$ is becoming large, the rotation is becoming very tiny. In some cases, an approximate QFT (AQFT) may be more accurate than a full QFT [23]. It is the same as overfitting data to a machine learning model. It is suggested to stop the recursion when $k$ reaches $log_2 n$. The cutoff reduces the complexity of transforms from $n(n+1)/2$ to roughly $n \cdot log_2 n$.

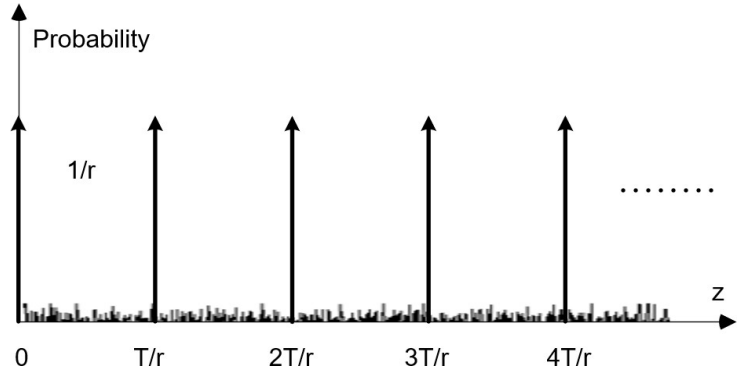

**Fig. 5.** Quantum Phase Additions in Frequency Domain

### 5.2 Analysis of Results

Quantum phase transforms are additions in frequency domain, The full recursion period-finding algorithm calls 196 time recursively to factor 35. For a number $N = 91(7 \times 13)$, the algorithm calls itself 257 times. The experiments were carried out on *ibmq_qasm_simulator* using *Ubuntu*18.10 *Cosmic*64 − *bit* virtual machine installed with *Qiskit*. Initially all 13 qubits are set to superposition states using the Hadamard gates. An additional two groups of registers one that contains 6 qubits and another one contains 9 qubits are used to store temporary results. A group of 14 quantum registers to store the final results after measurements. Using the AQFT method, the recursive calls were reduced to 54 and 178 respectively, a roughly $n \cdot log_2 n$ improvement without affecting the final results. In both cases, the quantum instance fires 1024 shots at the backend.

## 6 Conclusion

In this paper, we studied quantum algorithm performance improvement techniques in transpilation, optimization, error mitigation, and machine learning. An approximation-based novel optimization technique is proposed. The proposed AQFT algorithm improves performance in addition to exponential speedups for QFT computation.

## 7 Acknowledgment

This research is funded by National Science Foundation grants #2000136, #2329053, and a grant from USAF FA9550-25-1-0030.

## References

1. Laxmisha Rai, Smriti Khatiwada, Chunrao Deng, and Fasheng Liu. Cross-language code development with generative ai: A source-to-source translation perspective. In *2024 IEEE 7th International Conference on Electronic Information and Communication Technology (ICEICT)*, pages 562–565, 2024.
2. IBM. Ibm quantum computing platform. artificial intelligence qiskit code assistant, 2024.

3. Shuangbao Paul Wang and Eric Sakk. Quantum algorithms: Overviews, foundations, and speedups. In *IEEE 5th International Conference on Cryptography, Security and Privacy (CSP)*, pages 17–21, 2021.
4. Aniello Esposito, Jessica R. Jones, Sebastien Cabaniols, and David Brayford. A hybrid classical-quantum hpc workload. In *2023 IEEE International Conference on Quantum Computing and Engineering (QCE)*, volume 02, pages 117–121, 2023.
5. Qiskit. Quantum information software kit (qiskit). an open-source sdk for working with quantum computers at the level of extended quantum circuits, operators, and primitives, 2024.
6. Javier Robledo-Moreno, Mario Motta, Holger Haas, Ali Javadi-Abhari, Petar Jurcevic, William Kirby, Simon Martiel, Kunal Sharma, Sandeep Sharma, Tomonori Shirakawa, Iskandar Sitdikov, Rong-Yang Sun, Kevin J. Sung, Maika Takita, Minh C. Tran, Seiji Yunoki, and Antonio Mezzacapo. Chemistry beyond exact solutions on a quantum-centric supercomputer, 2024.
7. Sergiy Zhuk, Niall Robertson, and Sergey Bravyi. Trotter error bounds and dynamic multi-product formulas for hamiltonian simulation, 2024.
8. Niall F. Robertson, Bibek Pokharel, Bryce Fuller, Eric Switzer, Oles Shtanko, Mirko Amico, Adam Byrne, Andrea D'Urbano, Salome Hayes-Shuptar, Albert Akhriev, Nathan Keenan, Sergey Bravyi, and Sergiy Zhuk. Tensor network enhanced dynamic multiproduct formulas, 2024.
9. Jinjing Shi, Wenxuan Wang, Xiaoping Lou, Shichao Zhang, and Xuelong Li. Parameterized hamiltonian learning with quantum circuit. *IEEE Transactions on Pattern Analysis and Machine Intelligence*, 45(5):6086–6095, 2023.
10. Danil Kaliakin, Akhil Shajan, Javier Robledo Moreno, Zhen Li, Abhishek Mitra, Mario Motta, Caleb Johnson, Abdullah Ash Saki, Susanta Das, Iskandar Sitdikov, Antonio Mezzacapo, and Kenneth M. Merz Jr. Accurate quantum-centric simulations of supramolecular interactions, 2024.
11. Ieva Liepuoniute, Kirstin D. Doney, Javier Robledo-Moreno, Joshua A. Job, Will S. Friend, and Gavin O. Jones. Quantum-centric study of methylene singlet and triplet states, 2024.
12. Shuangbao (Paul) Wang. Design high-confidence computers using trusted instructional set architecture and emulators. *High-Confidence Computing*, 1(2):1–5, 2021.
13. Shuangbao Wang, Matthew Rohde, and Amjad Ali. Quantum cryptography and simulation: Tools and techniques. *ACM International Conference of Cryptography, Security and Privacy (ICCSP)*, pages 36 – 41, 2020.
14. Shuangbao Paul Wang. *Computer Architecture and Organization - Fundamentals and Architecture Security*. Springer, 2021.
15. Shuangbao Paul Wang and Robert S. Ledley. *Computer Architecture and Security*. Wiley, 2013.
16. IBM. Solve higher-order binary optimization problems with q-ctrl's optimization solver, 2024.
17. IBM. Dynamical decoupling, 2024.
18. Joel J. Wallman and Joseph Emerson. Noise tailoring for scalable quantum computation via randomized compiling. *Physical Review A*, 94(5), November 2016.
19. IBM Quantum. Combine error mitigation options with the estimator primitive., 2024.
20. Karl Mayer, Alex Hall, Thomas Gatterman, Si Khadir Halit, Kenny Lee, Justin Bohnet, Dan Gresh, Aaron Hankin, Kevin Gilmore, Justin Gerber, and John Gaebler. Theory of mirror benchmarking and demonstration on a quantum computer, 2023.
21. David Kremer, Victor Villar, Hanhee Paik, Ivan Duran, Ismael Faro, and Juan Cruz-Benito. Practical and efficient quantum circuit synthesis and transpiling with reinforcement learning, 2024.
22. Shuangbao Paul Wang and Paul A. Mullin. Trustworthy artificial intelligence for cyber threat analysis. In Kohei Arai, editor, *Intelligent Systems and Applications*, pages 493–504. Springer Lecture Notes, 2022.
23. Shuangbao (Paul) Wang. Recursion depth reduction in shor's algorithm. *Optica Publishing Group*, 2022.

## Authors

**Paul Wang** earned his Ph.D. in Computer Science advised by Dr. Robert Ledley, the inventor of the body CT scanner, at Georgetown University. Paul completed postdoc studies in Quantum Computing at MIT and in AI and Data Science at the University of Cambridge. Paul's research areas are quantum crypto, quantum networking, secure architecture, and trustworthy AI.

**Jianzhou Mao** received Ph.D. from Auburn University. He is currently a postdoctoral researcher in the Morgan quantum computing group.

**Eric Sakk** received his Ph.D. from Cornell University. He is an associate professor in the Morgan quantum computing group.